\documentclass[12pt]{article}

\usepackage{amsmath}
\usepackage{amssymb}

\numberwithin{equation}{section}

\begin{document}
\thispagestyle{empty}\setcounter{page}{0}

\begin{flushright}
UT-Komaba/05-8\\
hep-th/0508100\\
Aug, \ 2005
\end{flushright} 

\vspace{2mm}

\vspace{2ex}

\begin{center}
{\large \bf Comments on the Entropy and the Temperature of Non-extremal Black $p$-Brane }

\vspace{8ex}

{\large Ken-ichi Ohshima}

\vspace{3ex}
{\it  Institute of Physics, University of Tokyo\\
 Komaba, Meguro-ku 153 Tokyo }

\vspace{1ex}

email: ohshimak@hep1.c.u-tokyo.ac.jp\\ 

\end{center}

\vspace{6ex}

\centerline{Abstract} 

\begin{quote} 
We show that the exact entropy and the temperature (including coeffecient )
of non-extremal black $p$-brane are calculated by maximizing the entropy 
under several assumptions.  We argue the relation of those assumptions and
certain D$p$-D($p-2$) system in compactified space.
\end{quote}

\vspace{2ex}

\newpage

\vspace{4ex}

\section{\bf Introduction}
The microscopic origin of black hole entropy have been attached importance
for years since it is regarded that its mechanism will be explained in
quantum gravity theory. In string theory, Strominger and Vafa \cite{SV}
showed that the exact entropy of certain class of extremal black holes is actually obtained
from string theoretical microscopic model.  

     For  near extremal case, Gubser, Klebanov and Peet \cite{B3} showed that
D-brane and open string gas model provides the microscopic description of the Bekenstein-Hawking
entropy of black 3-brane. The open string degree of freedom is reduced by a factor of 3/4 in
their model, and this reduction is regarded as the strong coupling ( large quantum correction ) effect.
Soon \cite{Bp} showed that the supergravity result of the entropy of the near extremal $p$-branes in general
has simple relation  with the temperature and the charge like
\begin{equation}
    S \propto q^a T^b. \label{ff}
\end{equation}

    In 2001, Danielsson, Guijosa, Kruczenski \cite{DGK} showed that D-brane - open string gas model also
provide the microscopic description of  non-dilatonic Schwarzschild black branes and non-extremal
black branes.  The inovation they made is that the correct entropy up to coeffecient is
obtained by requiring the maximization of the entropy by the number of branes and antibranes.
\cite{SP} \cite{BL} extended it to general neutral $p$-brane and showed that similar method can be applied to 
get the correct entropy up to coeffecient.  It is also applied to non-extremal $p$-branes, muti-charged
branes, rotating branes \cite{Kal1} \cite{l} \cite{KalS} \cite{ght} \cite{Halyo}. 

      In these models for non-extremal case,  the energy of the gas on branes
 and antibranes must be indentical (to derive the correct entropy). This condition is mysterious in physical
 point of view especially in the fact that the temperature of the brane and antibrane are different
 and the Hawking temperature is described by $ 1/T_H = 1/T_{brane} + 1/T_{antibrane}$.
 
  In this paper, we show that the exact entropy and the temperature (including coeffecient) 
of non-extremal black $p$-brane are calculated  by maximizing the entropy 
under several assumptions.  For $p = 3$, the  exact entropy and the temperature are 
derived if we assume that the phenomena similar to \cite{LT} occur in 
black 3-brane or there are some correspondence between black 3-brane and D3-D1 system of \cite{LT}.
There is no "two temperatures" in our procedure so that the Hawking temperature is directly
calculated.  For general $p$, the exact entropy and the temperature are obtained from the similar procedure 
by requiring its statistical quantities should agree with supergravity result in near extremal limit.
 
 The organization of this paper is as follows. In section 2,  we point out some problems lying 
in this kind of arguments. In section 3, we show that the exact entropy 
and the temperature of black 3-brane are calculated, and argue  its relation to D3-D1 system in 
\cite{LT}. In section 4, we extend the
argument to general $p$-branes. Section 5 is the conclusion and discussion.

\section{ Problems lying in this kind of arguments }

 Before starting to explain our "prescription", we clarify several
fundamental problems lying in this kind of arguments.

The first problem is the non-BPS nature. Even for near extremal case,
large quauntum correction may exist at large $N$ limit, thus we
basically cannot utilize the quantities calculated at finte $N$ region for  
the quantities at large $N$ region. \cite{Mal} \cite{das} argued this 
problem  for near extremal case, but for general non-extremal
brane, this is open problem.  Even if one finds some model which leads to
the exact entropy or temperature or energy of non-extremal black $p$-branes, 
there is  no idea to justify it physically unless we have something like
"non-renormalisation theorem".
 
Even if we put aside the "non-renomalisation" problem and go forward to
find some corresponding D-brane model, we face second problem.
Many recent papers regarding  non-extremal black $p$-brane use 
the "maxmizing entropy" procedure ( this paper also use it ),  
but actually, for example for Schwarzschild black 3-brane,
such calculations lead to  (tension of branes) $\sim$ (gas energy)
as a result. This means
\begin{align}
N \tau_3 &\sim N^2 T^4  \ , \notag \\
&\downarrow \notag \\
 T^4 &\sim O( \dfrac{1}{ g_s N l_s^4 } ) \  , \notag
\end{align}
The temperature goes infinity at $g_s N << 1$ ; i.e.
we don't have low energy D-brane picture for Schwarzschild black 3-brane.
When $g_s N >> 1$, temperature is low but this time we cannot ignore 
the existence of horizon, because 
\begin{equation}
 r_H T = \dfrac{ 1 }{ \pi \cosh \gamma } \ , 
\end{equation}
where $r_H$ is horizon radius, and $\gamma \rightarrow 0$ for Schwarzschild limit;  i.e. 
we cannot ignore horizon radius in the energy scale of the Hawking temperature.
We must take the curved space and horizon into account, thus it 
cannot be described by D-brane model in flat space, even if we assume some
"non-renormalisation theorem".

 Moreover, when the magnitude of the gas energy is comparable to the 
tension of the branes, massless (free particle) gas model is quite 
mysterious.  Total energy of the statistical model is described as 
(massless gas) $+$  (brane tension), however it is difficult to consider 
it as "branes excited by gas",  because the excitation is too large to be described in 
Yang-Mills description.

 Taking into consideration the problems above, we start from some limit (but in total,
our argument contains imperfect step regarding this problem ).
Our model  has two parameters: one is the number of (charged) D3-branes $N$,
another is the number of (neutral charged) effective 3-brane $n$. Schwarzschild
case is $N = 0, \  n \neq 0$, and this limit contains the problem stated above.
Thus we start from the case of $g_s N <<1, \ g_s n <<1, \dfrac{n}{N} << g_s N $
(means slightly non-extremal ) where low energy D-brane picture is valid.
Then we extend it for $g_s N >>1, \ g_s n >>1$ to finite non-extremality and 
Schwarzschild brane. Although we start from low energy D-brane picture we needed 
to introduce small neutral mass of very high temperature in our argument,
and that is the imperfect step.

\section{ The entropy and temperature of  black 3-brane }

    We should point out first that the degree of freedom of open sting must
be reduced by 3/4 to derive exact near extremal black 3-brane entropy (and temperature) by 
ordinary D-brane - open string gas model\cite{B3}.
This 3/4 factor is regarded as strong coupling effect \cite{sia} \cite{atl}, but the detail
mechanism has not been  clarified yet.

     When starting to consider about non-extremal 3-branes, one faces another problem :
"How to add stable neutral charged mass to the system?".  Brane-antibrane system
is one candidate, but at least in flat space and in weak coupling limit and
 at low temperature, they are unstable and should be decay.  In \cite{DGK},
the stabilization of brane-antibrane system at large $N$ limit is argued.

     But "3/4 reduced d.o.f. open string gas" and "stable neutral mass" can be
simultaneously realized  at low temperature and at weak coupling  actually.
In \cite{LT}, by compactifying transverse 6 directions, stable neutral mass seems to be
realized as  "D1-vortices" in D3-brane.  RR 2-form field become massive and
D1 charges are screened.  Simultaneouly, gauge field on D3-brane also aquires mass.
It can be decoupled when the compactified volume $V_6 \rightarrow 0$, thus the open string
degree of freedom is reduced by 3/4. Now we have the model of stable neutral mass
and 3/4 reduced d.o.f. at low temperature successfully.

    Unfortunately the transverse direction of the black 3-brane that we consider is non-compact.
Thus the above \cite{LT} system seems to have nothing to do with the current problem, but
we will show below that the exact entropy and temperature is derived actually
not only for near extremal case but for arbitrary non-extremal case 
if we assume that the similar phenomena occur in the large $N$ limit or there are some
correspondences between black 3-brane and D3-D1 system of \cite{LT}.

\subsection{ Entropy and temperature by statistical model }

 First, open string gas model on the system \cite{LT} describes 
black 3-brane entropy and temperature at near extremal limit correctly . 
We have open string gas of 3/4 reduced d.o.f. and $N$ number of D3-branes in 
some limit of \cite{LT}, so the total energy 
is described as
\begin{equation}
E = N \tau_3 V \ + \ \dfrac{ 6 \pi }{ 16 } N^2 V T^4 \label{ne3} \ \ ,
\end{equation}
where the direction along the D3-brane is compactified as a torus, and $V$ is 
the volume of the torus. Above \eqref{ne3} leads to the exact black 3-brane 
entropy and temperature at near extremal limit, including coefficient \cite{B3}.
 
Next, we consider the situation that  "D1-vortices" exists inside D3-branes. 
According to \cite{LT}, the total brane mass seems to be described as
\begin{equation}
 \sqrt{ M_{D3}^2 + M_{D1}^2 } \label{tmass3} \ \ .
\end{equation}
\eqref{tmass3} resembles BPS bound of D3-D1, but note that at this time
total D1 charge can be neutral, due to massive RR 2-form field.     
We assume that total D1 charge is neutral and these D1 vortices form dense string network
in D3-branes so that the network can be described as effectively 3 dimensional branes.

Low energy approximation of fluctuation of 2-dimensional string network was argued in \cite{stn4}
and result is that it can be described as massless fields. In our case, the string network
is 3-dimensional and moreover turned into vortices in D3-branes, thus the situation 
is quite different.  But if we assume the fluctuation of the string network at low temperature
can be described as massless fields, the total energy is described as 
\begin{equation}
E = \sqrt{ N^2 + n^2 } \tau_3 V \ + \ f( N, n ) V T^4 \ \ ,\label{gn3} 
\end{equation}
where  $ f( N, n ) $ depends on the formation of the string network, 
and $n$ is the number of the "effective 3-branes" and can be continuous value.
The entropy of this system is
\begin{equation}
S = \dfrac{4}{3} f(N, n) V T^3 \ \ , \label{d13s}
\end{equation}
and we can rewrite it as
\begin{equation}
S = \dfrac{4}{3} f(N, n)^{1/4} V^{1/4} ( E -  \sqrt{ N^2 + n^2 } \tau_3 V )^{3/4} \ \ ,
\label{s32}
\end{equation}
from \eqref{gn3}. 

If $f(N, n)$ varies in the process before equilibrium, it varies 
with regard to $n$. Thus we miximize the entropy by $n$, then it leads to 
\begin{equation}
 ( E -  \sqrt{ N^2 + n^2 } \tau_3 V ) \dfrac{ \partial f}{\partial n}  = 3\dfrac{n}{ \sqrt{ N^2 + n^2} } \tau_3 V f \ .
\label{scond}
\end{equation}

Next we consider how to form the non-extremal black 3-brane.
Black hole can be described by few parameters, charge, mass and so on, 
as far as information is lost at the formation of the black hole.
In this point of view, the way of forming black hole is irrelevant, 
as far as the parameter of resulting black hole is the same.
We suppose the following case: one small (Schwarzschild) black 3-brane 
$m_s$ falls into one extremal black 3-brane (mass $M_e$ ) where
\begin{equation}
 M_e >> m_s \ ,
\end{equation}
so that this collision makes only small influence on the extremal black 3-brane.
(On the contrary, if $M_e \sim m_s$, it is difficult to predict the final state).

In the D-brane picture, the above situation means that small "neutral charged brane"
corresponding to Schwarzschild black brane,
\begin{equation}
E_s = n \tau_3 V \ + E_{\text{gas of neutral brane }} \ \ , \label{s3e} 
\end{equation}
gets close to $N$ number of coincident D-branes, 
\begin{equation}
E_e = N \tau_3 V \ .  \label{s3ex} 
\end{equation}

"Neutral charged brane" is realized by brane-antibrane system if 
the system has some mechanism of preventing tachyon condensation.
In \cite{DGK}, the stabilization of brane-antibrane system by 
finite temperature is argued. According to \cite{DGK}, 
the temperature
\begin{equation}
T  \ \  >> \ \  T_c \ \ \sim \ \dfrac{1}{ ( g_s n )^{1/2}  l_s } \ , \
\end{equation}
is needed to avoid tachyon condensation.  
However we are working at $g_s n << 1$ region, thus
the temperature of this neutral brane exceeds the Hagedorn temperature.  
For $n << 1$, the temperature is high enough to 
create D-brane so that the metastable state of creation / decay
of brane might be realized, but it is difficult to describe 
what is happening in this small neutral brane and 
also it is difficult to describe the physics before the equilibrium 
when this small neutral brane collides with the large number of extremal 
branes.

We assume that the temperature of the gas gets lower in 
sufficiently short time at the collision of this small neutral brane
and large number of extremal branes (of zero temperature). When the temperature
gets lower than  $T_c$,  the neutral brane has to decay to
lower dimensional branes such as D1-brane, and then it forms 
bound state with the extremal 3-brane. If this lowering occur
in sufficiently short time, most of the gas energy cannot 
turn into the tension of the neutral brane, thus the gas
energy will be almost conserved after equilibrium:
\begin{equation}
E_{\text{gas of non-extremal}} \ \simeq \ E_{\text{gas of neural brane }} \ .
\end{equation}
$E_{\text{gas of neural brane }}$ can be written as
\begin{equation}
E_{\text{gas of neural brane }} \ = \ \beta \ n \tau_3 V \ ,
\end{equation}
of some $\beta$, then we can write as
\begin{equation}
  E -  \sqrt{ N^2 + n^2 } \tau_3 V   = \beta \ n \tau_3 V \ ,
\end{equation}
and plugging this into \eqref{scond}, we get the equation
\begin{equation}
  \dfrac{ \partial f}{\partial n}  = \dfrac{ 3 / \beta }{ \sqrt{ N^2 + n^2} }  f \ .
\end{equation}
The solution is
\begin{equation}
 f(N, n ) = C ( \sqrt{N^2 + n^2} + n )^{3/\beta} \ ,
\end{equation}
where $C$ is a constant. We have the following statistical model
for non-extremal 3-brane now :
\begin{align}
E &= \sqrt{ N^2 + n^2 } \tau_3 V  \ + \  C ( \sqrt{N^2 + n^2} + n )^{3/\beta} \  V T^4 \label{d13e2} \ , \\
S &= \dfrac{4}{3}  C ( \sqrt{N^2 + n^2} + n )^{3/\beta} \  V T^3 \label{d13s2} \ . 
\end{align}
We can take the near extremal limit ($\dfrac{n}{N} \rightarrow 0$) of this model smoothly. The 
near extremal limit of \eqref{d13e2} \eqref{d13s2} is
\begin{align}
E &= N \tau_3 V +  C N^{3/\beta} \   V T^4 \label{d13e_ne} \ , \\
S &= \dfrac{4}{3}  C N^{3/\beta} \   V T^3 \label{d13s_ne} \ .
\end{align} 
We already know the near extremal limit as \eqref{ne3}. By comparing
\eqref{d13e_ne} and \eqref{ne3} we get
\begin{align}
C &= \dfrac{6 \pi^2}{ 16 } \ , \\ 
\beta &= \dfrac{ 3 }{ 2 } \ \label{betae} \ .
\end{align}
Then the statistical model of the non-extremal 3-brane is decided as
\begin{align}
E &= \sqrt{ N^2 + n^2 } \tau_3 V  \ + \  \dfrac{6 \pi^2}{ 16 } ( \sqrt{N^2 + n^2} + n )^2  V T^4 \label{sb3e}\\
S &= \dfrac{ \pi^2}{ 2 } ( \sqrt{N^2 + n^2} + n )^2  V T^3 \label{sb3s} \ .
\end{align}
The entropy and the temperature are written as
\begin{align}
S &= 2^{1/2} \pi^{1/2} ( \sqrt{N^2 + n^2} + n )^{1/2} n^{3/4} \tau_3^{3/4} V \label{fn3s} \ , \\
T &= \dfrac{ 2^{1/2} \tau_3^{1/4} n^{1/4} }{ \pi^{1/2} ( \sqrt{N^2 + n^2} + n )^{1/2} } \ . \label{fn3t}
\end{align}

\subsection{Comparing with supergravity result }

The mass(total energy), charge per unit volume , entropy and temperature of  black 3-brane by supergravity 
is \cite{Bp}\cite{adm}\cite{bbd}\cite{spb},
\begin{align}
E &= \dfrac{\pi^3}{ 2 \kappa^2 } V \mu^4 ( 5 \ + \ 4 \sinh^2 \gamma ) \ , \\
q &= \dfrac{ 2 \pi^3}{  \sqrt{2} \kappa } \mu^4 \sinh 2 \gamma \ , \\
S &= \dfrac{ 2 \pi^4 }{ \kappa^2 } \mu^5 V \cosh \gamma \ , \\
T &= \dfrac { 1 }{ \pi \mu \cosh \gamma } \ ,
\end{align} 
where $\mu$ and $\gamma$ are parameters.

The above mass, entorpy and temperature can be written by the number of 
D3-branes as 
\begin{align}
E &= \sqrt{ N^2 + n^2 } \tau_3 V + \dfrac{ 3 }{ 2 } n \tau_3 V \ , \\
S &= 2^{1/2} \pi^{1/2} ( \sqrt{N^2 + n^2} + n )^{1/2} n^{3/4} \tau_3^{3/4} V \ , \\
T &= \dfrac{ 2^{1/2} \tau_3^{1/4} n^{1/4} }{ \pi^{1/2} ( \sqrt{N^2 + n^2} + n )^{1/2} } \ , 
\end{align} 
and those exactly agree with  \eqref{fn3s} \eqref{fn3t} \eqref{betae} derived from the statistical
model, {\it at all the point from near extremal limit to Schwarzschild case}.

We have started from the region $g_s N <<1, \ g_s n <<1, \dfrac{n}{N} << g_s N $ in
the D-brane model of previous subsection, but above shows the entropy and the temperature
match exactly in the region $g_s N >>1, \ g_s n >>1, \dfrac{n}{N} \text{: arbitrary} $.
It looks like the correspondence of the region $g_s N <<1, \ g_s n <<1, \dfrac{n}{N} << g_s N $
and the region of $g_s N >>1, \ g_s n >>1$,  where $N$ and $n$ can be 
made large independently.

\subsection{ Neutral mass as  string network }

We have assumed that neutral charged mass forms dense string network inside 
D3-branes. But so far the neutral charged mass are dealt like $n \tau_3 V $
as if it were "neutral charged 3-brane". If the D3-D1 system of \cite{LT} 
with D1-vortices are related to black 3-brane and our assumption is correct, 
there must be some evidences of string network.

Actually the energy of the gas has the following relation with the Hawking temperature:
\begin{equation}
  E_{gas} = \dfrac{ 6 \pi^2 }{ 16 } N^2 V T_H^4 \ +  \ \dfrac{3 \pi}{2} n^{3/2} \tau_3^{1/2} V T_H^2 \ .
\end{equation}
The first term is for the gas on the (charged) $N$ number of D3-branes. 
The second term implies the excitation of one dimensional object, and moreover,  
depends only neutral mass $n$ and does not depend on $N$.
This implies that after the equilibrium, the gas on the neutral mass moves along some
 1 dimensional objects.

\section{ The entropy and temperature of  black p-brane }

We can easily extend the argument in the previous section to general $p$-brane.
However, we have no near extremal statistical model ( like D-brane and open string
gas model as the previous section ), because $p$-brane(except $p=3$) in 10 dimension is 
dilatonic in general.  Thus we present the procedure to derive the exact entropy  
and temperature of non-extremal black $p$-brane from its near extremal result of 
supergravity. 

\subsection{ Supergravity result }

The mass(total energy), charge per unit volume, entropy and temperature of the single charged 
black p-brane of 1/2 BPS is \cite{Bp}\cite{adm}\cite{bbd}\cite{spb} , 

\begin{align}
E &= \dfrac{ \omega_{d+1} }{ 2  \kappa^2 } \mu^d  V ( d + 1 \ + \ d \sinh^2 \gamma ) \ ,\\
q &= \dfrac{ \omega_{d+1} }{ 2 \sqrt{2 }  \kappa } d \mu^d \sinh 2\gamma \ ,\\
S &= \dfrac{ 2 \pi \omega_{d+1} }{ \kappa^2 } \mu^{d+1} V \cosh \gamma \ ,\\
T &= \dfrac{d}{4 \pi \mu \cosh \gamma } \ , \label{ptempe}
\end{align}
where $d = 7 - p$, $\omega_{d+1}$ is the volume of a $d+1$-dimensional unit sphere,
$V$ is the volume of the torus which the brane wrapped, $\mu$ and $\gamma$ are parameters.  

We rewrite the mass and the entropy in terms of the number of (charged ) branes $N$
\begin{align}
E &= \sqrt{ N^2 + n^2 } \tau_p V \ + \ 2 \lambda n \tau_p V \ , \label{pmassn} \\ 
S &= B V \ (\sqrt{ N^2 + n^2 } + n )^{1/2} n^\lambda \label{pentn} \ ,
\end{align}
where 
\begin{align}
 n  &= \dfrac{\omega_{d+1} \ d \ \mu^d }{ 4 \kappa^2 \tau_p} \ , \\
 B &= 2^{2\lambda + 3/2} \pi \ \omega_{d+1}^{1/2 - \lambda} d^{-(\lambda+1/2)}  \kappa^{2\lambda -1}  \tau_p^{\lambda+1/2} \ . 
\end{align}
As in \cite{Bp}, the entropy and the temperature have simple relation as 
\eqref{ff} at near extremal limit( $n/N \ll 1 $ ).  The total energy in 
near extremal limit is
\begin{equation}
E =  N \tau_p V \ + \ (2 \tau_p)^{\frac{-\lambda}{1-\lambda}} B^{\frac{1}{1-\lambda}} \lambda  \ N^{\frac{1}{2(1-\lambda)} } V T^{\frac{1}{1-\lambda}}  \ .
\label{pne}
\end{equation}
The first term corresponds to the tension of the branes and the second term corresponds to
"gas".

\subsection{ Entropy and temperature from statistical model }

As in 3-brane case, we assume D$p$ - D($p-2$) bound state model like
\begin{equation}
E =  \sqrt{ N^2 + n^2 } \tau_p V \ +  \ f( N, n ) V T^\alpha \ \ . \label{pe1} 
\end{equation}
This time we left the power of $T$ as unknown parameter $\alpha$.
Comparing \eqref{pe1} at near extremal limit with \eqref{pne}, the parameter $\alpha$
is decided as
\begin{equation}
\alpha = \dfrac{1}{1-\lambda}  \ . \
\end{equation}
The entropy of this system is
\begin{align}
S &= \dfrac{1}{\lambda} \   f( N, n ) V T^ { \frac{\lambda}{1-\lambda} } \\
 &=   \dfrac{1}{\lambda} \   f^{1-\lambda} V^{1-\lambda} ( E -  \sqrt{ N^2 + n^2 } \tau_p V )^\lambda \label{pent2} \ .
\end{align}
We maximize the entropy by $n$  as in previous section,
\begin{equation}
 (1 - \lambda) ( E -  \sqrt{ N^2 + n^2 } \tau_p V ) \dfrac{ \partial f}{\partial n}  = \dfrac{ \lambda n \tau_p V}{ \sqrt{ N^2 + n^2} } f \ .
\label{pscond}
\end{equation}
As in previous section, we can write the gas energy as
\begin{equation}
   E -  \sqrt{ N^2 + n^2 } \tau_p V  \  =  \ \beta n \tau_p V \ .
\end{equation}
Plugging this into \eqref{pscond}, we get
\begin{equation}
 \dfrac{ \partial f}{\partial n} = \dfrac{ \lambda}{ \beta ( 1-\lambda)} \dfrac{1}{ \sqrt{ N^2 + n^2} } f \ .
\end{equation}
The solution is 
\begin{equation}
 f = C ( \sqrt{ N^2 + n^2 } + n )^{ \frac{ \lambda}{ \beta ( 1-\lambda)} } \ ,
\end{equation}
then the total energy is
\begin{equation}
E =  \sqrt{ N^2 + n^2 } \tau_p V  \ +     C ( \sqrt{ N^2 + n^2 } + n )^{ \frac{ \lambda}{ \beta ( 1-\lambda)} } \ V \ T^{ \frac{1}{1-\lambda} } \ .
\end{equation}
In the near extremal limit, 
\begin{equation}
E = N \tau_p V  \ + \ C N^{ \frac{ \lambda}{ \beta ( 1-\lambda)} } \ V \ T^{ \frac{1}{1-\lambda} } \ .
\end{equation}
Comparing above with \eqref{pne}, unknown constant $C$ and $\beta$ are decided as
\begin{align}
\beta &= 2 \ \lambda \ , \\
C &= (2 \tau_p )^{\frac{-\lambda}{1-\lambda}} B^{\frac{1}{1-\lambda}} \lambda \ .
\end{align}
Substitute this to \eqref{pent2}, we get entropy 
\begin{equation}
S = B V  (\sqrt{ N^2 + n^2 } + n )^{1/2} n^\lambda \ ,
\end{equation}
and this agrees perfectly with supergravity result \eqref{pentn}. Also, temperature exactly agrees with 
\eqref{ptempe}.

\section{Conclusion and discussion}

In this paper, we showed that exact entropy and temperature of non-extremal black 3-brane
are calculated by maxmizing entropy under several assumptions. These assumptions  
imply the correspondence of black 3-brane statistical mechanics and certain D3-D1 system. 
We extend this argument to general $p$-branes and presented a procedure to derive
exact entropy and temperature of non-extremal black $p$-brane by requiring its 
statistical quantities to agree with supergravity result at near extremal limit.

As written in section 2, even if we confirm that the entropy or other thermodynamical
quantities are derived from some model in perturbative regime, still it is 
mysterious unless we have something like non-renormalization theorem.   

But the perfect agreement of entropy, temperature, degree of freedom of gas
and the total tension of the brane may be nontrivial.  Moreover, the system becomes
unstable when RR charge = 0, because the screening of D($p-2$) charge breaks when
D$p$ branes do not exist( the screening caused by the coupling of the 
RR field and the gauge field on D$p$ branes \cite{LT}).  This may be related to 
Gregory-Laflamme instability of black $p$-branes \cite{gl1} \cite{gl2}.  We will further 
investigate on the correspondence between the D$p$ \ - \ D$(p-2)$ model and 
black $p$-brane.

\vspace{5ex}

{\bf Acknowledgements}\\
 I would like to thank to Tamiaki Yoneya, Mitsuhiro Kato, Koji Hashimoto, Kenji Hotta, Yuri Aisaka 
for many helpful discussions.

\end{document}